\documentclass[12pt]{article}
\usepackage{amsmath,amssymb}
\usepackage{graphicx}
\allowdisplaybreaks[1]

\makeatletter
 
  \@addtoreset{equation}{section}
 \makeatother
\newcommand{\bfe}{{\mathbf e}}
\newcommand{\bfk}{{\mathbf k}}
\newcommand{\bfl}{{\mathbf l}}

\newcommand{\bfq}{{\mathbf q}}
\newcommand{\bfr}{{\mathbf r}}

\newcommand{\bm}{\bar{m}}
\newcommand{\bn}{\bar{n}}
\newcommand{\bz}{\bar{z}}

\newcommand{\bphi}{{\bar{\phi}}}
\newcommand{\bpsi}{\bar{\psi}}

\newcommand{\bxi}{\bar{\xi}}

\newcommand{\bsigma}{\bar{\sigma}}

\newcommand{\bareta}{\bar{\eta}}
\newcommand{\blambda}{\bar{\lambda}}

\newcommand{\hd}{\hat{d}}
\newcommand{\hm}{\hat{m}}
\newcommand{\hn}{\hat{n}}
\newcommand{\hz}{\hat{z}}

\newcommand{\hmu}{\hat{\mu}}
\newcommand{\hlambda}{\hat{\lambda}}

\newcommand{\tphi}{\tilde{\phi}}
\newcommand{\tpsi}{\tilde{\psi}}

\newcommand{\txi}{\tilde{\xi}}

\newcommand{\cD}{{\cal D}}
\newcommand{\cF}{{\cal F}}
\newcommand{\cG}{{\cal G}}

\newcommand{\cL}{{\cal L}}
\newcommand{\cN}{{\cal N}}

\newcommand{\dalpha}{{\dot{\alpha}}}

\newcommand{\tbpsi}{{\bar{\tilde{\psi}}}}
\newcommand{\tbphi}{{\bar{\tilde{\phi}}}}

\newcommand{\bhz}{\bar{\hat{z}}}

\newcommand{\Z}{\mathbb{Z}}

\newcommand{\nn}{\nonumber}

\newcommand{\Tr}{{\rm Tr}\,}
\newcommand{\del}{\partial}

\newcommand{\ul}{\underline}

\begin{document}

%%
%% TITLE
%%
\begin{titlepage}

%% Set the number of the title with 0
\setcounter{page}{0}
\renewcommand{\thefootnote}{\fnsymbol{footnote}}

\vspace{20mm}

\begin{center}
{\large\bf
Two-dimensional $\cN=(2,2)$
Supersymmetric Lattice Gauge Theory 
with Matter Fields in the Fundamental Representation
}

\vspace{20mm}
{
So Matsuura\footnote{{\tt matsuura@nbi.dk}}}\\
\vspace{10mm}

{\em The Niels Bohr International Academy, 
The Niels Bohr Institute, \\
Blegdamsvej 17, DK-2100 Copenhagen, Denmark}

\end{center}

\vspace{20mm}
\centerline{{\bf Abstract}}
\vspace{10mm}
In this paper, we construct a lattice formulation for 
two-dimensional $\cN=(2,2)$ supersymmetric gauge theory 
with matter fields in the fundamental representation.  
We first construct it by the orbifolding procedure from 
Yang-Mills matrix theory with eight supercharges. 
We show that we can obtain the same lattice formulation by extending 
the geometrical discretization scheme. 
This suggests that the equivalence between the two schemes 
holds even for theories with matter fields.

\end{titlepage}
\newpage

\renewcommand{\thefootnote}{\arabic{footnote}}
\setcounter{footnote}{0}

\section{Introduction}

Supersymmetric gauge theory is one of the most exciting
topics in high energy physics from various points of view. 
Among many attempts to understand the nature of supersymmetric 
gauge theory, 
there have recently been many important developments towards 
putting exactly preserved supersymmetries on a 
space-time lattice. 
In %Kaplan
\cite{Cohen:2003xe}%
--\nocite{Cohen:2003qw}\cite{Kaplan:2005ta}, 
several lattice formulations have been constructed 
by the so-called orbifolding procedure 
from Yang-Mills matrix theories.%
\footnote{ 
{}For further analysis, 
see, $e.g.$, refs. \cite{Catterall:2006jw}%
--\nocite{Giedt:2003ve}\nocite{Onogi:2005cz}%
\nocite{Ohta:2006qz}\cite{Damgaard:2007be}.}
In %Catterall
\cite{Catterall:2003wd}--%
\nocite{Catterall:2004np}\nocite{Catterall:2005fd}\cite{Catterall:2007kn}, 
Catterall formulated several lattice theories 
using a general prescription, 
the so-called geometrical discretization. 
In %Sugino
\cite{Sugino:2003yb}--\nocite{Sugino:2004qd}%
\nocite{Sugino:2004uv}\cite{Sugino:2006uf}, 
Sugino discretized topologically twisted 
gauge theories while keeping the BRST symmetry on a lattice. 
One characteristic feature of Sugino's lattice formulations is that 
link variables are expressed by unitary matrices 
like conventional lattice gauge theories, 
which is a desirable condition 
for numerical simulations 
\cite{Suzuki:2007jt}\cite{Kanamori:2007ye}\cite{Kanamori:2007yx}. 
In %Kawamoto
\cite{D'Adda:2004jb}--\nocite{D'Adda:2005zk}%
\nocite{D'Adda:2007ax}\cite{Nagata:2008zz}, 
the authors constructed lattice theories
based on a deformed supersymmetry 
algebra on a lattice.% 
\footnote{
For a discussion on consistency in the deformation of 
supersymmetry on a lattice, 
see \cite{Bruckmann:2006ub}\cite{Bruckmann:2006kb}%
\cite{Arianos:2007nv}\cite{D'Adda:2007ax}. 
See also \cite{Nagata:2008xk} 
for a further discussion on the consistency 
connecting with large-N limit.}
See \cite{Unsal:2005us} for a different approach
to lattice supersymmetry in terms of a deformed type IIB 
matrix model without orbifolding. 
For other approaches to examine supersymmetric gauge theories
on a lattice, see 
%Others
\cite{Nishimura:1997vg}--\nocite{Kaplan:1983sk}\nocite{Maru:1997kh}%
\nocite{Neuberger:1997bg}\nocite{Kaplan:1999jn}\nocite{Fleming:2000fa}%
\nocite{Montvay:2001aj}%
\nocite{Suzuki:2005dx}%
\nocite{Elliott:2005bd}\cite{Elliott:2007bt}.
For numerical approach to supersymmetric theories 
without using lattice formulations, 
see \cite{Hanada:2007ti}\cite{Anagnostopoulos:2007fw}.%
\footnote{
See also 
\cite{Kawahara:2007fn}--\nocite{Kawahara:2007ib}%
\nocite{Catterall:2007fp}\cite{Catterall:2008yz}
for numerical study of black hole thermodynamics
and gauge/gravity duality. 
}

One of the most important recent results 
in supersymmetric lattice gauge theory is that 
the above seemingly different lattice formulations 
which preserve supersymmetry on a lattice 
are related to each other.
Indeed, the geometrical discretization scheme 
was found to be equivalent with the orbifolding procedure
\cite{Damgaard:2007xi}\cite{Catterall:2007kn}%
\cite{Damgaard:2008pa}. 
We can directly derive the prescription of the geometrical 
discretization scheme by combining a dimensional reduction 
and the orbifolding procedure. 
This means that the geometrical discretization 
gives an effective shortcut to the orbifolding procedure. 
Practically, this equivalence makes it easy to identify 
the naive continuum limit of a lattice theory 
since, using this equivalence, 
we can directly construct the lattice formulation 
from the continuum theory. 
In \cite{Takimi:2007nn}, Sugino's lattice formulation 
was shown to be derived 
from Catterall's complexified lattice theory \cite{Catterall:2004np} 
by restricting the degrees of freedom of 
the complexified fields while preserving the supercharge. 
{}Furthermore, in \cite{Damgaard:2007eh}, 
the formulations provided by the link approach 
were also shown to be the same with those of orbifolding.

The purpose of this paper is to construct a lattice theory for 
two-dimensional $\cN=(2,2)$ supersymmetric gauge theory 
with matter fields in the fundamental representation
using both the schemes of the orbifolding 
procedure and the geometrical discretization. 
In \cite{Endres:2006ic}, the authors constructed a lattice 
theory for two-dimensional $\cN=(2,2)$ supersymmetric gauge theory 
with matter fields in the {\em adjoint} representation  
and suggested that 
matter fields in the fundamental representation 
can be introduced by 
considering an additional $Z_2$ transformation 
in the orbifolding procedure. 
In this paper, we explicitly realize this idea to construct
a lattice theory for  
two-dimensional $\cN=(2,2)$ supersymmetric gauge theory 
coupled with matter fields in the fundamental representation 
from a mother theory with eight supercharges.%
\footnote{
For an alternative application of this idea, see \cite{Giedt:2006dd}.
} 
We derive the same lattice formulation from the continuum 
theory by slightly extending the geometrical discretization scheme 
so that we can apply it to a theory with matter fields. 

The organization of this paper is as follows.
In the next section, we construct a lattice formulation 
for two-dimensional $\cN=(2,2)$ supersymmetric gauge theory 
with hypermultiplets in the fundamental representation 
using the orbifolding procedure. 
In \S 3, we extend the prescription of the geometrical discretization 
scheme and derive the same lattice action that is constructed in \S 2. 
\S 4 is devoted to the conclusion and some discussions.

\section{Construction of lattice theory via orbifolding procedure}

In this section, we construct a lattice theory of two-dimensional 
$\cN=(2,2)$ supersymmetric gauge theory coupled with hypermultiplets 
in the fundamental representation 
using the orbifolding procedure. 
We start with a matrix theory (mother theory) with 
eight supercharges 
used in \cite{Cohen:2003qw}:
\begin{align}
S_{\rm m} = \Tr\Bigl(
&\frac{1}{4}|[z_a,z_b]|^2 
+\frac{1}{2}[z_a,\bz_a] D 
-\frac{1}{2} D^2 \nn  \\
&+ \eta[\bar{z}_a,\psi_a]
+ \frac{1}{2}\xi_{ab}\left(
[z_a,\psi_b]-[z_b,\psi_a]
\right)
+ \frac{1}{2}\chi_{abc}[\bar{z}_a,\xi_{bc}]
\Bigr), 
\label{mother action for 8 SUSY}
\end{align}
where $a,b,c=1,2,3$, $z_a$ and $\bz_a$ are bosonic 
complex matrices,
$D$ is a bosonic Hermitian matrix (an auxiliary field),
and $\eta$, $\psi_a$, $\xi_{ab}$ and $\chi_{abc}$ are fermionic 
complex matrices, which are assumed to be antisymmetric under 
the permutation of the indices. 
We also assume that all the matrices are of the size $(N_c+N_f)N^2$. 
This action is invariant under a global symmetry $SO(6)\times SU(2)$
and a gauge symmetry $U((N_c+N_f)N^2)$. 
This is obvious from the fact that (\ref{mother action for 8 SUSY}) 
is obtained by dimensionally reducing the action of 
six-dimensional $\cN=1$ supersymmetric Yang-Mills theory 
followed by an appropriate field redefinition \cite{Cohen:2003qw} 
(see also \cite{Damgaard:2007be}). 
For the purpose of later discussion we rewrite 
$\{z_3,\bz_3,\psi_3,\xi_{m3},\chi_{123}\}$ 
as 
$\{\phi,\bar\phi,\bareta,\bpsi_m,\bxi_{12}\}$. 
Then the action  
(\ref{mother action for 8 SUSY}) can be rewritten as 
\begin{align}
 S_{\rm m} = \Tr \Bigl(
&\frac{1}{4}\left|[z_m,z_n]\right|^2 
+\frac{1}{2}\left|[z_m,\phi]\right|^2
+ \frac{1}{2}\left([z_m,\bz_m]+[\phi,\bar\phi]\right) D 
-\frac{1}{2}D^2\nn \\
&+\eta[\bz_m,\psi_m] 
+ \frac{1}{2}\xi_{mn}\left([z_m,\psi_n]-[z_n,\psi_m]\right) \nn \\
&+\bareta[z_m,\bpsi_m] 
+ \frac{1}{2}\bxi_{mn}\left([\bz_m,\bpsi_n]-[\bz_n,\bpsi_m]\right) \nn \\
&+\eta[\bar\phi,\bareta]+\bpsi_m[\phi,\psi_m]
+\frac{1}{2}\bxi_{mn}[\bar\phi,\xi_{mn}]
\Bigr), 
\label{mother action 2 for 8 SUSY}
\end{align}
where $m,n=1,2$. 
In the orbifolding procedure, the maximal $U(1)$ symmetry, 
$U(1)^4$ in this case, plays an important role. 
In the expression (\ref{mother action for 8 SUSY}) 
or (\ref{mother action 2 for 8 SUSY}), the $U(1)$
symmetry is manifest and the charge assignment is as follows:
\begin{center}
\begin{tabular}{c|cccccccccccc}
 & $z_1$ & $z_2$ & $z_3$ & D & $\eta$ & $\xi_{23}$ & $\xi_{31}$ & $\xi_{12}$ 
 & $\chi_{123}$ & $\psi_1$ & $\psi_2$ & $\psi_3$  \\
\hline
$q_1$ & 1 & 0 & 0 & 0 & 1/2 & 1/2 & -1/2 & -1/2 & 1/2 & 1/2 & -1/2 & -1/2 \\
$q_2$ & 0 & 1 & 0 & 0 & 1/2 & -1/2 & 1/2 & -1/2 & 1/2 & -1/2 & 1/2 & -1/2 \\
$q_3$ & 0 & 0 & 1 & 0 & 1/2 & -1/2 & -1/2 & 1/2 & 1/2 & -1/2 & -1/2 & 1/2 \\
$q_4$ & 0 & 0 & 0 & 0 & 1/2 & 1/2 & 1/2 & 1/2 & -1/2 & -1/2 & -1/2 & -1/2 
\end{tabular}
\end{center}
In order to construct a two-dimensional lattice theory with at least 
one preserved supercharge, we define two different 
charges 
$(r_1,r_2)$ as two different 
linear combinations of $q_i$ with requiring  
$r_1=r_2=0$ for $\eta$ \cite{Damgaard:2007be}. 
In addition to this $U(1)^2$ symmetry, we further consider 
a $Z_2$ symmetry \cite{Endres:2006ic} that transforms 
the fields 
as $\Phi\to e^{s \pi i} \Phi$ $(s=0,1)$ 
corresponding to ``parity'' associated with the fields. 
We define $s$ as
\begin{equation}
 s\equiv q_3 - q_4 +1 \quad ({\rm mod}\  2). 
\end{equation}
As a result, 
the $U(1)$ charges and the parity are summarized as 
\begin{center}
\begin{tabular}{c|ccccccccccc}
& $z_m$ & $\bz_m$ & $\phi$ & $\bar\phi$ & $D$ & $\eta$ & $\psi_m$ & $\xi_{12}$ 
& $\bareta$ & $\bpsi_m$ & $\bxi_{12}$ \\ \hline
$(r_1,r_2)$ & $\bfe_m$ & $-\bfe_m$ & $\bfq$ & $-\bfq$ & $0$ & $0$ 
& $\bfe_m$ & $-\bfe_1-\bfe_2$ & $\bfq$ & $-\bfe_m-\bfq$ &
					     $\bfe_1+\bfe_2+\bfq$ \\
 $s$ & $0$ & $0$ & $1$ & $1$ & $0$ & $0$ & $0$ & $0$ & $1$ & $1$ & $1$ 
\end{tabular}
\end{center}
where $\bfe_1$ and $\bfe_2$ are two linearly independent integer valued 
two-vectors and $\bfq$ is a linear combination of $\bfe_1$ and
$\bfe_2$. 
As shown in \cite{Damgaard:2007be}, the lattice structure is determined 
not by the detail of the assignment of the $U(1)$ 
charges but the linear relation
between $\bfe_m$ and $\bfq$. Therefore we assume 
$\bfe_m=\hat{m}$; a unit vector in the positive $m$'th direction 
in the following.

Using the $U(1)$ charges and the parity given above, 
we carry out the following two kinds of 
orbifold projections. 
We first define two $Z_N^2$ transformations generated 
by 
\begin{equation}
 \gamma_i : \Phi \to \omega^{r_i} \Omega_i \Phi \Omega_i^{-1}, \quad
  (i=1,2)
\label{Z_N trans.}
\end{equation}
where $\omega=e^{2\pi i/N}$, $r_i$ are $U(1)$ charges 
of a matrix $\Phi$ 
and $\Omega_i$ are defined by 
\begin{align}
 \Omega_1&= \Bigl(
{\mathbf 1}_{N_c} \otimes U_N \otimes {\mathbf 1}_N 
\Bigr)
\oplus 
\Bigl(
 {\mathbf 1}_{N_f} \otimes U_N \otimes {\mathbf 1}_N\Bigr),  \nn \\
 \Omega_2&= \Bigl(
{\mathbf 1}_{N_c} \otimes {\mathbf 1}_N \otimes U_N \Bigr) \oplus
\Bigl(
{\mathbf 1}_{N_f} \otimes {\mathbf 1}_N \otimes U_N\Bigr) ,   
\end{align}
using the clock matrix, 
$U_N\equiv{\rm diag}(\omega,\omega^2,\cdots,\omega^N)$.
In addition,
%following a comment in \cite{Endres:2006ic}, 
we define a $Z_2$ transformation generated by 
\begin{equation}
 p: \Phi \to e^{s\pi i} P \Phi P, \qquad
P\equiv \left(\begin{matrix}
{\mathbf 1}_{N_cN^2} & {\mathbf 0}  \\
{\mathbf 0} & -{\mathbf 1}_{N_fN^2}
\end{matrix}\right),
\label{Z_2 trans.}
\end{equation}
where $e^{s \pi i}$ is the parity associated 
with the matrix $\Phi$.

The orbifold projection is defined by projecting out 
such elements of each matrix that are not invariant under the
transformations (\ref{Z_N trans.}) and (\ref{Z_2 trans.}). 
To explain how these projections work, 
we express a general 
matrix $\Phi$ by four blocks of the size 
$N_cN^2\times N_c N^2$, 
$N_cN^2\times N_f N^2$, $N_fN^2\times N_c N^2$ 
and $N_fN^2\times N_fN^2$ as 
\begin{equation}
 \Phi=\left(\begin{matrix}
\Phi_{11} & \Phi_{12}  \\
\Phi_{21} & \Phi_{22}  
\end{matrix}\right). 
\end{equation}
Suppose the $U(1)$ charges of $\Phi$ is $\bfr=(r_1,r_2)$. 
Then, after the projection associated with (\ref{Z_N trans.}), each
block is expressed as 
\begin{equation}
 \Phi_{ij} = \sum_{\bfk\in\Z_N^2} \Phi_{ij}(\bfk) \otimes
  E_{\bfk,\bfk+\bfr}, \quad (i,j=1,2) 
\end{equation}
where $E_{\bfk,\bfl}=E_{k_1,l_1}\otimes E_{k_2,l_2}$ with 
$\left(E_{kl}\right)_{mn}=\delta_{km}\delta_{ln}$
and 
$\Phi_{ij}(\bfk)$ is a matrix with the size $N_c\times N_c$, 
$N_c\times N_f$, $N_f\times N_c$ and $N_f\times N_f$ corresponding 
to $(i,j)=(1,1), (1,2), (2,1)\ {\rm and}\ (2,2)$, respectively.
Furthermore, if $\Phi$ is parity even(odd), 
the blocks $\Phi_{12}$ and $\Phi_{21}$ ($\Phi_{11}$ and $\Phi_{22}$)
are projected out by (\ref{Z_2 trans.}). 
As a result, after the projections (\ref{Z_N trans.})
and (\ref{Z_2 trans.}), $\Phi$ with $s=0$ can be written as 
\begin{align}
 \Phi^{(s=0)} &=
\left(\begin{matrix}
\sum_{\bfk} \Phi_{11}(\bfk)\otimes E_{\bfk,\bfk+\bfr} & {\mathbf 0}\\
{\mathbf 0} & \sum_{\bfk} \Phi_{22}(\bfk)\otimes E_{\bfk,\bfk+\bfr} 
\end{matrix}\right),
\label{s=0}
\end{align}
and $\Phi$ with $s=1$ can be written as 
\begin{align}
\Phi^{(s=1)}&=
\left(\begin{matrix}
{\mathbf 0} & \sum_{\bfk} \Phi_{12}(\bfk)\otimes E_{\bfk,\bfk+\bfr}  \\
\sum_{\bfk} \Phi_{21}(\bfk)\otimes E_{\bfk,\bfk+\bfr} & {\mathbf 0}
\end{matrix}\right).
\label{s=1}
\end{align}

We introduce a notation to express (\ref{s=0}) 
and (\ref{s=1}) as 
\begin{equation}
\Phi^{(s=0)}= \sum_{\bfk} \left(
\Phi_{11}(\bfk)\otimes E_{\bfk,\bfk+\bfr} +
\Phi_{22}(\bfk)\otimes E_{N+\bfk,N+\bfk+\bfr}
\right), 
\label{s=0-2}
\end{equation}
and
\begin{equation}
\Phi^{(s=1)} =  \sum_{\bfk} \left(
\Phi_{12}(\bfk)\otimes E_{\bfk,N+\bfk+\bfr} 
+ \Phi_{21}(\bfk)\otimes E_{N+\bfk,\bfk+\bfr}
\right), 
\label{s=1-2}
\end{equation}
respectively. 
Then, from the assignment of the $U(1)$ charges and 
the parity given above, we see that 
the matrices $\{z_m,\bz_m,D,\eta,\psi_m,\xi_{12}\}$ 
can be written as 
\begin{align}
 z_m&\equiv\sum_{\bfk\in\Z_N^2} \left(
z_m(\bfk)\otimes E_{\bfk,\bfk+\hm}
+\hz_m(\bfk)\otimes E_{N+\bfk,N+\bfk+\hm} 
\right), \nn \\
\bz_m&\equiv\sum_{\bfk\in\Z_N^2} \left(
\bz_m(\bfk)\otimes E_{\bfk+\hm,\bfk}
+\bhz_m(\bfk)\otimes E_{N+\bfk+\hm,N+\bfk}\right), \nn \\
D &\equiv \sum_{\bfk\in\Z_N^2} \left(
d(\bfk)\otimes E_{\bfk,\bfk} + 
\hat{d}(\bfk)\otimes E_{N+\bfk,N+\bfk}
\right), \nn \\
\eta &\equiv \sum_{\bfk\in\Z_N^2} \left(
\lambda(\bfk)\otimes E_{\bfk,\bfk} + 
\hlambda(\bfk)\otimes E_{N+\bfk,N+\bfk}
\right), \nn \\
\psi_m &\equiv \sum_{\bfk\in\Z_N^2} \left(
\lambda_m(\bfk)\otimes E_{\bfk,\bfk+\hm} +
\hlambda_{m}(\bfk) \otimes E_{N+\bfk,N+\bfk+\hm}
\right), \nn \\
\xi_{mn}&\equiv \sum_{\bfk\in\Z_N^2} \left(
\lambda_{mn}(\bfk)\otimes E_{\bfk+\hm+\hn,\bfk} + 
\hlambda_{mn}(\bfk)\otimes E_{N+\bfk+\hm+\hn,N+\bfk}
\right),
\label{after projection 1}
\end{align}
and the others can be written as 
\begin{align}
\phi &\equiv \sum_{\bfk\in\Z_N^2} \left(
-\sqrt{2}i \tbphi(\bfk)\otimes E_{\bfk,N+\bfk+\bfq}
+\sqrt{2}i \bphi(\bfk+\bfq) \otimes E_{N+\bfk,\bfk+\bfq}
\right), \nn \\
\bphi &\equiv \sum_{\bfk\in\Z_N^2} \left(
-\sqrt{2}i \phi(\bfk+\bfq)\otimes E_{\bfk+\bfq,N+\bfk}
+\sqrt{2}i \tphi(\bfk)\otimes E_{N+\bfk+\bfq,\bfk}
\right), \nn \\
\bareta &\equiv \sum_{\bfk\in\Z_N^2} \left(
\tbpsi(\bfk)\otimes E_{\bfk,N+\bfk+\bfq}
+\bpsi(\bfk+\bfq)\otimes E_{N+\bfk,\bfk+\bfq}
\right), \nn \\
\bpsi_m&\equiv \sum_{\bfk\in\Z_N^2} \left(
\psi_m(\bfk+\hm+\bfq)\otimes E_{\bfk+\hm+\bfq,N+\bfk}
+\tpsi_m(\bfk)\otimes E_{N+\bfk+\hm+\bfq,\bfk}
\right), \nn \\
\bxi_{mn}&\equiv \sum_{\bfk\in\Z_N^2} \left(
\tbpsi_{mn}(\bfk)\otimes E_{\bfk,N+\bfk+\hm+\hn+\bfq}
+\bpsi_{mn}(\bfk+\hm+\hn+\bfq)\otimes E_{N+\bfk,\bfk+\hm+\hn+\bfq}
\right). 
\label{after projection 2}
\end{align}
The action of the orbifold lattice theory is obtained 
by substituting (\ref{after projection 1}) 
and (\ref{after projection 2}) into the action 
of the mother theory (\ref{mother action 2 for 8 SUSY}) 
and expand it around a classical configuration, 
\begin{align}
 z_m(\bfk)=\bz_m(\bfk)= \frac{1}{a} {\mathbf 1}_{N_c}, 
\quad 
 \hz_m(\bfk)=\bhz_m(\bfk)= \frac{1}{a} {\mathbf 1}_{N_f}.  
\end{align}
Note that the obtained theory preserves only 
one supercharge for any value of $\bfq$. 
This can be seen by the fact that
only one of the eight supersymmetry parameters 
in the mother theory 
has the charge assignment $\bfr=0$ and $s=0$. 
Since the supersymmetry parameters are c-numbers, 
the other seven supersymmetry parameters are 
projected out by the orbifold projections \cite{Damgaard:2007eh}. 
The preserved supersymmetry transformation on the lattice 
is obtained by substituting (\ref{after projection 1}) 
and (\ref{after projection 2}) into the supersymmetry 
transformation of the matrices in the mother theory:
\begin{align}
 Q z_m &= \psi_m, \quad 
 Q \bz_m = 0, \quad 
 Q D = [\psi_m,\bz_m] + [\bareta,\bphi], \nn \\
\label{SUSY algebra}
 Q \eta &= \frac{1}{2}\left([z_m,\bz_m]+[\phi,\bphi]-D \right), \quad 
 Q \psi_m = 0, \quad 
 Q \xi_{mn} = -\frac{1}{2}[\bz_m,\bz_n], \\
 Q \phi &= \bareta, \quad 
 Q \bphi = 0, \quad
 Q \bareta = 0, \quad
 Q \bpsi_m = -\frac{1}{2}[\bz_m,\bphi], \quad
 Q \bxi_{mn} = 0, \nn 
\end{align}
followed by the shift, $z_m \to 1/a + z_m$ and 
$\bz_m \to 1/a + \bz_m$.

\begin{figure}[t]
  \begin{center}
    \includegraphics[scale=.6]{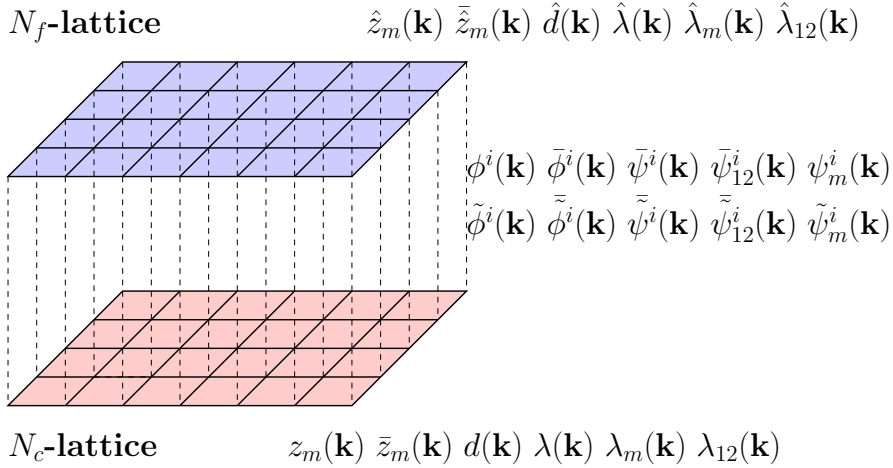}
  \end{center}
  \caption{
 The lattice space-time obtained by the two kinds of 
orbifold projections corresponding to (\ref{Z_N trans.})
and (\ref{Z_2 trans.}). 
The lattice variables 
$\{z_m(\bfk),\,\allowbreak\bz_m(\bfk),\,
\allowbreak d(\bfk),\,
\allowbreak\lambda(\bfk),\,
\allowbreak\lambda_m(\bfk),\,
\allowbreak\lambda_{12}(\bfk)\}$
live on the $N_c$-lattice and 
$\{\hz_m(\bfk),\,\allowbreak \bar{\hz}_m(\bfk),\, 
\hat{d}(\bfk),\,\allowbreak\hlambda(\bfk),\,
\hlambda_m(\bfk),\,\allowbreak 
\hlambda_{12}(\bfk)\}$ 
live on the $N_f$-lattice.
The matter fields 
$\{\phi^i(\bfk),\,
\allowbreak\bar{\tilde{\phi^i}}(\bfk),\,
\allowbreak\bar{\tilde{\psi^i}}(\bfk),\,
\allowbreak\bar{\tilde{\psi^i}}_{12}(\bfk),\, 
\allowbreak\psi_m^i(\bfk)\}$ and 
$\{\bphi^i(\bfk),\,
\allowbreak\tilde{\psi^i}(\bfk),\,
\allowbreak\tilde{\psi^i}(\bfk),\,
\allowbreak\tilde{\psi^i}_{12}(\bfk),\,
\allowbreak\bpsi_m^i(\bfk)\}$ 
live on links connecting the two lattice space-times. 
 }
  \label{lattice-fig}
\end{figure}
From the construction, 
the obtained theory is a supersymmetric 
lattice gauge theory 
with a gauge group $U(N_c)\times U(N_f)$ 
defined on two copies of two-dimensional 
lattice space-times of size $N^2$ 
(Fig.~\ref{lattice-fig}).  
The lattice variables 
$\{z_m(\bfk),\,\allowbreak\bz_m(\bfk),\,
\allowbreak d(\bfk),\,
\allowbreak\lambda(\bfk),\,
\allowbreak\lambda_m(\bfk),\,
\allowbreak\lambda_{12}(\bfk)\}$
and  
$\{\hz_m(\bfk),\,\allowbreak \bar{\hz}_m(\bfk),\, 
\hat{d}(\bfk),\,\allowbreak\hlambda(\bfk),\,
\hlambda_m(\bfk),\,\allowbreak 
\hlambda_{12}(\bfk)\}$ 
transform in the representation 
$({\rm adj},\,{\mathbf 1})$ 
and $({\mathbf 1},\,{\rm adj})$ 
of $U(N_c) \times U(N_f)$, 
respectively, 
and live on different lattice space-times. 
In the following, we call these two lattice space-times 
the $N_c$-lattice 
and the $N_f$-lattice, respectively. 
On the other hand, 
$\{\phi^i(\bfk),\,
\allowbreak\bar{\tilde{\phi^i}}(\bfk),\,
\allowbreak\bar{\tilde{\psi^i}}(\bfk),\,
\allowbreak\bar{\tilde{\psi^i}}_{12}(\bfk),\, 
\allowbreak\psi_m^i(\bfk)\}$ and 
$\{\bphi^i(\bfk),\,
\allowbreak\tilde{\psi^i}(\bfk),\,
\allowbreak\tilde{\psi^i}(\bfk),\,
\allowbreak\tilde{\psi^i}_{12}(\bfk),\,
\allowbreak\bpsi_m^i(\bfk)\}$ 
transform in the representation  
$(\square,\overline{\square})$ and 
$(\overline{\square},\square)$ 
of $U(N_c)\times U(N_f)$, respectively, 
and live on links connecting 
the two lattice space-times.  
Although this lattice action for this quiver gauge theory 
is not our main purpose, it is an important result. 
We write down the action in appendix A.

Finally, in order to construct a lattice theory 
with matter fields in the fundamental representation 
of the gauge group $U(N_c)$, 
we make the fields living on the $N_f$-lattice non-dynamical 
by hand:
\begin{align}
 \hz_m(\bfk)&=\hat{\bz}_m(\bfk)
 =\hlambda(\bfk)=\hlambda_m(\bfk)=\hlambda_{12}(\bfk)=0, \nn \\
 \hd(\bfk)&=\bphi(\bfk+\bfq)\phi(\bfk+\bfq)
-\tphi(\bfk-\bfq)\tbphi(\bfk-\bfq).
\label{freeze}
\end{align}
The restricted theory is still supersymmetric 
since this restriction does not conflict with 
the supersymmetry (\ref{SUSY algebra}). 
By this operation, 
the symmetry $U(N_f)$ is no longer 
a gauge symmetry but a flavor symmetry 
and we obtain a lattice action with $U(N_c)$ gauge 
symmetry. 
After integrating out the auxiliary field $d(\bfk)$, 
we obtain, 
\begin{equation}
 S_{\rm lat} = 
S_{\rm gauge} + S_{\rm matter}, 
\label{lattice action}
\end{equation}
with
\begin{align}
 S_{\rm gauge}=\Tr_{N_c}\sum_{\bfk}\Bigl(
&\frac{1}{4}\left| 
\nabla_m^+ z_n(\bfk) - \nabla_n^+ z_m(\bfk)
+z_m(\bfk)z_n(\bfk+\hm) - z_n(\bfk)z_m(\bfk+\hn)
%\cF_{mn}(\bfk)
\right|^2 \nn \\
&+\frac{1}{8}
\left(
\nabla_m^- \left(z_m(\bfk)+\bz_m(\bfk)\right)
+z_m(\bfk)\bz_m(\bfk)-\bz_m(\bfk-\hm) z_m(\bfk-\hm)\right)^2
\nn \\
&-\lambda(\bfk)\bar\cD_m^-\lambda_m(\bfk)
+\frac{1}{2}\lambda_{mn}(\bfk)\left(
\cD_m^+\lambda_n(\bfk)-\cD_n^+\lambda_m(\bfk)
\right) 
\Bigr)
\label{gauge lattice action}
\end{align}
\begin{align}
 S_{\rm matter}=\sum_{\bfk}\Bigl(
&\frac{1}{2} \bar{\cD}_m^+ \bphi^i(\bfk) \cD_m^+ \phi^i(\bfk)  
+\frac{1}{2} {\cD}_m^+ \bphi^i(\bfk) \bar\cD_m^+ \phi^i(\bfk) \nn \\
&+\frac{1}{2} \bar{\cD}_m^+ \tphi^i(\bfk) \cD_m^+ \tbphi^i(\bfk)  
+\frac{1}{2} {\cD}_m^+ \tphi^i(\bfk) \bar\cD_m^+ \tbphi^i(\bfk) \nn \\
&+\frac{1}{4}\Tr_{N_c}\left( 
\tbphi^i(\bfk)\tphi^i(\bfk)
-\phi^i(\bfk)\bphi^i(\bfk)
\right)^2 \nn \\
%%%%%%
&+\bpsi^{i}(\bfk)
\cD_m^-
\psi_m^{i}(\bfk+\hm) 
+\tpsi_m^{i}(\bfk)
\cD_m^-
\bar{\tilde{\psi^i}}(\bfk+\hm) \nn \\
&-\frac{1}{2}\bpsi_{mn}^{i}(\bfk+\hm+\hn)
  \bigl[
\bar{\cD}_m^+
\psi_n^{i}(\bfk+\hn)
-\bar{\cD}_n^+
\psi_m^{i}(\bfk+\hm)
  \bigr] \nn \\
&-\frac{1}{2}
 \bigl[
  \tpsi_n^{i}(\bfk+\hm)
\bar{\cD}_m^+
\bar{\tilde{\psi^i}}_{mn}(\bfk)
  -\tpsi_m^{i}(\bfk+\hn)
\bar{\cD}_n^+
\bar{\tilde{\psi^i}}_{mn}(\bfk))
 \bigr] \nn \\
 &+\sqrt{2}i\bigl(
  \bpsi^{i}(\bfk)\lambda(\bfk)\phi^{i}(\bfk)
  -\tphi^{i}(\bfk)\lambda(\bfk)\bar{\tilde{\psi^i}}(\bfk) \bigr) \nn\\
 &+\sqrt{2}i\bigl(
 -\tpsi_m^{i}(\bfk)\lambda_m(\bfk)\bar{\tilde{\phi^i}}(\bfk+\hm)
  +\bphi^{i}(\bfk)\lambda_m(\bfk)\psi_m^{i}(\bfk+\hm)\bigr) \nn \\
 &+\frac{\sqrt{2}i}{2}\bigl(
 \bpsi_{mn}^{i}(\bfk+\hm+\hn)\lambda_{mn}(\bfk)\phi^{i}(\bfk)
 -\tphi^{i}(\bfk+\hm+\hn)\lambda_{mn}(\bfk)\bar{\tilde{\psi^i}}_{mn}(\bfk)
 \bigr)\Bigr), 
\label{matter lattice action}
\end{align}
where 
$i=1,\cdots,N_f$ is a flavor index, 
$\cD_m^{\pm}$ and $\bar{\cD}_m^{\pm}$ 
are covariant difference defined as 
\begin{align}
{\cD}_m^+ \Phi_{\rm adj}(\bfk) &\equiv 
 \nabla_m^+ \Phi_{\rm adj}(\bfk) 
 +z_m(\bfk)\Phi_{\rm adj}(\bfk+\hm)
 -\Phi_{\rm adj}(\bfk)z_m(\bfk+\bfr), \nn \\
\bar{\cD}_m^+ \Phi_{\rm adj}(\bfk) &\equiv 
 \nabla_m^+ \Phi_{\rm adj}(\bfk)
  +\Phi_{\rm adj}(\bfk+\hm)\bz_m(\bfk+\bfr)
  -\bz_m(\bfk)\Phi_{\rm adj}(\bfk),  \nn \\
{\cD}_m^- \Phi_{\rm adj}(\bfk) 
&\equiv \cD_m^+ \Phi_{\rm adj}(\bfk-\hm), \quad
\bar{\cD}_m^- \Phi_{\rm adj}(\bfk) 
\equiv \bar{\cD}_m^+ \Phi_{\rm adj}(\bfk-\hm), 
\end{align}
for a lattice field in the adjoint representation living on a link 
$(\bfk,\bfk+\bfr)$, 
\begin{align}
{\cD}_m^+ \Phi_{\square}(\bfk) &\equiv 
 \nabla_m^+ \Phi_{\square}(\bfk) 
 +z_m(\bfk)\Phi_{\square}(\bfk+\hm), \nn \\
\bar{\cD}_m^+ \Phi_{\square}(\bfk) &\equiv 
 \nabla_m^+ \Phi_{\square}(\bfk) 
  -\bz_m(\bfk)\Phi_{\square}(\bfk),  \nn \\
{\cD}_m^- \Phi_{\square}(\bfk) 
&\equiv \cD_m^+ \Phi_{\square}(\bfk-\hm), \quad
\bar{\cD}_m^- \Phi_{\square}(\bfk) 
\equiv \bar{\cD}_m^+ \Phi_{\square}(\bfk-\hm), 
\end{align}
for a lattice field in the fundamental representation, and 
\begin{align}
{\cD}_m^+ \Phi_{\overline{\square}}(\bfk) &\equiv 
 \nabla_m^+ \Phi_{\overline{\square}}(\bfk) 
  -\Phi_{\overline{\square}}(\bfk)z_m(\bfk), \nn \\
\bar{\cD}_m^+ \Phi_{\overline{\square}}(\bfk) &\equiv 
 \nabla_m^+ \Phi_{\overline{\square}}(\bfk) 
 +\Phi_{\overline{\square}}(\bfk+\hm)\hz_m(\bfk)
 \nn \\
{\cD}_m^- \Phi_{\overline{\square}}(\bfk) 
&\equiv \cD_m^+ \Phi_{\overline{\square}}(\bfk-\hm), \quad
\bar{\cD}_m^- \Phi_{\overline{\square}}(\bfk) 
\equiv \bar{\cD}_m^+ \Phi_{\overline{\square}}(\bfk-\hm), 
\end{align}
for a lattice field in the anti-fundamental representation. 
Here, 
$\nabla_m^{\pm}$ are forward and backward differences 
defined by 
\begin{align}
 \nabla_m^+ f(\bfk)
&=\frac{1}{a}\left(f(\bfk+\hm)-f(\bfk)\right),\quad 
 \nabla_m^- f(\bfk)=\frac{1}{a}\left(f(\bfk)
-f(\bfk-\hm)\right). 
\end{align}

The gauge part of the action 
(\ref{gauge lattice action}) is nothing but 
the lattice action for two-dimensional 
$\cN=(2,2)$ supersymmetric Yang-Mills theory given 
in \cite{Cohen:2003xe}. 
On the other hand, as we see in the next section, 
the continuum limit of (\ref{matter lattice action}) 
is the action of hypermultiplets of two-dimensional 
$\cN=(2,2)$ supersymmetry.  
Using the language of four-dimensional $\cN=1$ supersymmetry, 
it is obtained by a dimensional reduction to two dimensions 
from the action, 
\begin{equation}
 S_{\rm 4D}^{\rm matter} 
= \int d^4 x d^2\theta d^2 \bar{\theta} \Bigl(
\bar{\Phi}^i e^{2V} \Phi^i 
+\tilde{\Phi}^i e^{-2V} \bar{\tilde{\Phi}}
\Bigr),
\label{4D matter}
\end{equation}
where $\Phi^i$ and $\tilde{\Phi}^i$ are 
four-dimensional $\cN=1$ chiral superfields 
in the fundamental and anti-fundamental representations, 
respectively.
Therefore, the action (\ref{lattice action}) gives 
a supersymmetric lattice formulation 
for two-dimensional $\cN=(2,2)$ 
supersymmetric gauge theory coupled with 
hypermultiplets in the fundamental representation. 
The correspondence between lattice variables 
and continuum fields becomes clear 
in the next section.
Note that the matter fermions in this lattice theory 
do not have doublers. 
In fact, the fermion determinant is proportional to 
$\nabla_m^+ \nabla_m^-$, 
which has zero only at
the origin of the momentum space.

We close this section by making two comments. 
First, the matter fields in this lattice theory 
live only on sites even if they have non-zero 
$U(1)$ charges in general. 
Although it seems peculiar at first sight, 
we can understand it by seeing that we have two 
lattice space-times and matter fields live on links 
between the $N_c$-lattice and the $N_f$-lattice.
Since the $N_f$-lattice
becomes invisible
by the operation (\ref{freeze}), the matter fields 
behave as site variables on the $N_c$-lattice.

%Correspondingly, 
%the chiral symmetry is explicitly broken in this lattice theory. 
%In the continuum theory, the axial $U(1)$ transformation 
%rotates not only the fermions but also the adjoint scalar fields. 
%However, as discussed in \cite{Cohen:2003xe}, 
%the gauge field $A_m$ and two (real) adjoint
%scalar fields $\varphi_1$
%and $\varphi_2$ 
%in the continuum theory are combined into 
%complex link variables 
%on a lattice as 
%\begin{equation}
%z_m=\varphi_m + i A_m.
%\label{boson correspondence}
%\end{equation} 
%Since the orbifold projections for $z_1$ and $z_2$ are different, 
%the chiral $U(1)$ symmetry is explicitly broken on a lattice.

Second, we can consistently truncate the matter fields 
with tilde
from the action (\ref{matter lattice action}). 
This corresponds to the configuration, 
\begin{align}
 \phi &\equiv \sum_{\bfk\in\Z_N^2} \left(
\sqrt{2}i \bphi(\bfk+\bfq) \otimes E_{N+\bfk,\bfk+\bfq}
\right), \quad
\bphi \equiv \sum_{\bfk\in\Z_N^2} \left(
-\sqrt{2}i \phi(\bfk+\bfq)\otimes E_{\bfk+\bfq,N+\bfk}
\right), \nn \\
\bareta &\equiv \sum_{\bfk\in\Z_N^2} \left(
\bpsi(\bfk+\bfq)\otimes E_{N+\bfk,\bfk+\bfq}
\right), \quad
\bpsi_m\equiv \sum_{\bfk\in\Z_N^2} \left(
\psi_m(\bfk+\hm+\bfq)\otimes E_{\bfk+\hm+\bfq,N+\bfk}
\right), \nn \\
\bxi_{mn}&\equiv \sum_{\bfk\in\Z_N^2} \left(
\bpsi_{mn}(\bfk+\hm+\hn+\bfq)\otimes E_{N+\bfk,\bfk+\hm+\hn+\bfq}
\right),
\label{trancation}
\end{align}
instead of (\ref{after projection 2}).
In the language of four-dimensional $N=1$ theory, 
this is a dimensionally reduced theory of 
\begin{equation}
\cL=\int d^2\theta d^2 
\bar{\theta}\bar{\Phi}^i e^{2V} \Phi^i.
\end{equation}

\section{Derivation via geometrical discretization}

In this section, we derive the same lattice action 
(\ref{lattice action}) by extending 
the geometrical 
discretization scheme constructed by Catterall. 

In the geometrical discretization scheme, 
we construct a lattice theory from a continuum theory. 
Since it is known that the gauge part of the 
lattice action (\ref{gauge lattice action}) can 
be obtained 
by the geometrical discretization scheme 
\cite{Catterall:2007kn}, we concentrate on the 
matter action (\ref{matter lattice action}). 
In order to write down the continuum action, 
we start with the action 
of four-dimensional Euclidean 
$\cN=1$ theory (\ref{4D matter}), which is 
written down in component as  
\begin{align}
 \cL_{\rm 4D}^{{\rm matter}}=&
D_\mu\bphi^i(x) D_\mu \phi^i(x) 
+ D_\mu \tphi^i(x)D_\mu\bar{\tilde{\phi^i}}(x) 
+ \bphi^i(x) d(x) \phi^i(x) 
- \tphi^i(x) d(x) \bar{\tilde{\phi^i}}(x) \nn \\
&-i \bxi^i(x) \bsigma^\mu D_\mu \xi^i(x) 
-i \txi^i(x) \sigma^\mu D_\mu \bar{\tilde{\xi^i}}(x)\nn \\
&+i\sqrt{2}\left(\bphi^i(x)\lambda(x)\xi^i(x)
-\bxi^i(x)\blambda(x)\phi^i(x)\right) \nn \\
&+i\sqrt{2}\left(\tphi^i(x)\blambda(x)\bar{\tilde{\xi^i}}(x)
-\txi^i(x)\lambda\bar{\tilde{\phi^i}}(x)\right),
\label{4D component action} 
\end{align}
where $\mu=1,\cdots,4$, 
$D_\mu$ is a four-dimensional 
covariant derivative,
$\{\phi^i(x),\tbphi^i(x)\}$ and 
$\{\bphi^i(x),\tilde{\phi^i}(x)\}$ are 
complex scalar fields in the fundamental 
and anti-fundamental representations, respectively, 
$\lambda(x)$ and $\blambda(x)$ are two-component 
spinors in adjoint representation, 
$\xi^i(x)$ and $\bar{\tilde{\xi^i}}(x)$ 
are two-component spinors 
in the fundamental representation, 
and $\bxi^i(x)$ and $\txi^i(x)$ 
are two-component spinors in 
the anti-fundamental representation.% 
\footnote{The notation is based on \cite{Wess-Bagger}.}

We next dimensionally reduce (\ref{4D component action})
to two dimensions spanned by $\{x^4,x^2\}$. 
Correspondingly, 
we rename $\{v_4,v_2\}$ and $\{v_3,v_1\}$ as 
$\{A_1,A_2\}$ and $\{\varphi_1,\varphi_2\}$, 
respectively, 
and write the component of the spinors as 
\begin{align}
 \lambda_\alpha(x)&=\left(\begin{matrix}
\lambda_1(x) \\ \lambda_2(x)
\end{matrix}\right),\ 
\blambda^{\dot{\alpha}}(x)=\left(\begin{matrix}
\lambda_{12}(x) \\ -\lambda(x)
\end{matrix}\right),\ 
\xi^i_\alpha(x)=\left(\begin{matrix}
-\psi_2^i(x) \\ \psi_1^i(x)
\end{matrix}\right),\nn \\ 
\bxi^{i\dot{\alpha}}(x)&=\left(\begin{matrix}
\bpsi^i(x) \\ -\bpsi^i_{12}(x)
\end{matrix}\right),\
\txi_\alpha^i(x) =  \left(\begin{matrix}
-\tpsi_2^i(x) \\ \psi_1^i(x)
\end{matrix}\right),\
\bar{\tilde{\xi}}^{i \dalpha}(x) 
= \left(\begin{matrix}
\bar{\tilde{\psi^i}}(x) \\ -\bar{\tilde{\psi^i}}_{12}(x)
\end{matrix}\right). 
\label{fermion correspondence}
\end{align}
Then, after integrating out the auxiliary field, 
we obtain the following expression of 
two-dimensional matter action:
\begin{align}
 S_{\rm matter} = 
\int d^2x \Bigl(
&\frac{1}{2} \cD_m \bphi^{i}(x) \bar{\cD}_m \phi^{i}(x)
+\frac{1}{2} \bar\cD_m \bphi^{i}(x) {\cD}_m \phi^{i}(x) \nn \\
&+\frac{1}{2} \cD_m\tphi^{i}(x) \bar{\cD}_m \bar{\tilde{\phi^i}}(x) 
+\frac{1}{2} \bar\cD_m \tphi^{i}(x) {\cD}_m \bar{\tilde{\phi^i}}(x)
\nn \\
&+\frac{1}{4}\Tr_{N_c}\left(
\phi^{i}(x)\bphi^{i}(x)
-\bar{\tilde{\phi^i}}(x) \tphi^{i}(x)
\right)^2\nn \\
&+\bpsi^{i}(x)\cD_m\psi_m^{i}(x) 
+\tpsi^{i}(x)\cD_m\bar{\tilde{\psi^i_m}}(x) \nn \\ 
&-\frac{1}{2}\bigl[
 \bpsi_{mn}^{i}(x)\bar{\cD}_m \psi_n^{i}(x)
   -\bpsi_{mn}^{i}(x)\bar{\cD}_n \psi_m^{i}(x)
  \bigr] \nn \\
&-\frac{1}{2}
 \bigl[
  \tpsi_n^{i}(x) \bar{\cD}_m 
\bar{\tilde{\psi^i}}_{mn}(x)
  -\tpsi_m^{i}(x) \bar{\cD}_n \bar{\tilde{\psi^i}}_{mn}(x)
 \bigr] \nn \\
 &+\sqrt{2}i\bigl(
  \bpsi^{i}(x)\lambda(x)\phi^{i}(x)  
  -\tphi^{i}(x)\lambda(x)\bar{\tilde{\psi^i}}(x)\bigr) \nn \\
 &+\sqrt{2}i\bigl(
  -\tpsi_m^{i}(x)\lambda_m(x)\bar{\tilde{\phi^i}}(x)
  +\bphi^{i}(x)\lambda_m(x)\psi_m^{i}(x)\bigr) \nn \\
 &+\frac{\sqrt{2}i}{2}\bigl(
 \bpsi_{mn}^{i}(x)\lambda_{mn}(x)\phi^{i}(x)
  -\tphi^{i}(x)\lambda_{mn}(x)\bar{\tilde{\psi^i}}_{mn}(x)
 \bigr) 
\Bigr), 
\label{2D continuum action}
\end{align}
where we have defined 
$\cD_m = \del_m +i A_m + \phi_m$ and 
$\bar{\cD}_m = \del_m +i A_m - \phi_m$ \cite{Catterall:2007kn}. 

The prescription of the geometrical discretization 
scheme given in \cite{Catterall:2004np} 
is summarized as the following four rules:
\begin{enumerate}
 \item{An adjoint $p$-form field is mapped to 
a lattice variable on a $p$-cell. The gauge 
transformation for a $p$-form field 
$f_{\mu_1\cdots\mu_p}(\bfk)$ is given by
$f_{\mu_1\cdots\mu_p}(\bfk)\to 
g(\bfk)f_{\mu_1\cdots\mu_p}(\bfk)
g^{-1}(\bfk+\hmu_1+\cdots+\hmu_p)$
or
$f_{\mu_1\cdots\mu_p}(\bfk)\to 
g(\bfk+\hmu_1+\cdots+\hmu_p)
f_{\mu_1\cdots\mu_p}(\bfk)
g^{-1}(\bfk)$ depending on the direction 
of the $p$-cell.%
\footnote{The direction of the $p$-cell is 
determined by the $U(1)$ charge of the $p$-form.
For detail, see \cite{Damgaard:2008pa}. 
}}
 \item{A curl-like covariant differential 
is mapped to a covariant forward difference.}
 \item{A divergent-like covariant differential
is mapped to a covariant backward difference.}
\item{An interaction term is written so that 
it forms a loop on a lattice.}
\end{enumerate}
Since this scheme is originally constructed for a theory 
that contains only adjoint fields,  
we add the following rule for 
fields in the fundamental representation:
\begin{enumerate}
 \item[5.]{A field in the fundamental or anti-fundamental 
representation is mapped on a site. The gauge 
transformation is $\psi(\bfk)\to g(\bfk)\psi(\bfk)$
and $\bpsi(\bfk)\to\bpsi(\bfk)g^{-1}(\bfk)$
for fields in the fundamental and anti-fundamental 
representations, respectively.} 
\end{enumerate}
Applying these rules to (\ref{2D continuum action}),
it is easy to see that we obtain the matter 
part of the lattice action (\ref{matter lattice action}). 
Combining the result in \cite{Catterall:2007kn}, 
we conclude that the equivalence between 
the orbifolding procedure and the geometrical 
discretization scheme still holds for a theory 
with matter fields in the fundamental representation. 
In this scheme, the correspondence between 
lattice variables and continuum fields is manifest. 
Thus we can conclude that the continuum limit of the lattice 
theory given by (\ref{lattice action}) is two-dimensional 
$\cN=(2,2)$ supersymmetric gauge theory coupled with 
hypermultiplets.

\section{Conclusion and discussion}

In this paper, we have constructed  
a lattice formulation of two-dimensional $\cN=(2,2)$
supersymmetric $U(N_c)$ gauge theory coupled with 
hypermultiplets in the fundamental representation. 
We have constructed the theory using the orbifolding 
procedure from the mother theory with eight supercharges 
by combining two kinds of orbifold projections. 
The obtained lattice theory preserves only one supercharge. 
We have also shown that the same lattice action 
can be constructed by extending 
the geometrical discretization scheme. 
This suggests that the equivalence between 
these two schemes to construct a supersymmetric 
lattice theory holds even for a theory with 
matter fields.

In the construction of the model (\ref{lattice action}), 
we started with a mother theory 
with eight supercharges while the 
the obtained theory is a lattice theory for 
two-dimensional $\cN=(2,2)$ 
supersymmetric gauge theory, which has four supercharges. 
This is the first example in the orbifolding procedure 
where the number of supercharges of the mother 
theory and that of the obtained (continuum) theory 
is different.  
In fact, all the lattice formulations constructed so far 
are those for theories with the same number of supercharges 
with the mother theory 
\cite{Cohen:2003xe}\cite{Cohen:2003qw}\cite{Kaplan:2005ta}%
\cite{Endres:2006ic}\cite{Damgaard:2007be}. 
Although 
it is anticipated that the additional orbifold projection 
corresponding to the $Z_2$ transformation (\ref{Z_2 trans.})
would be the origin of this phenomenon, 
the reason is still unclear. 
Moreover, at present, 
there is no principle to use this $Z_2$ transformation 
to obtain an $\cN=(2,2)$ theory. 
Even if we start with the same mother theory, 
we can construct several different lattice formulations with fields 
in the fundamental representation by changing the definition 
of the $Z_2$ transformation. 
It would be an interesting future work to clarify 
a general principle to construct a lattice formulation 
with matter fields that has a proper continuum limit.

Another interesting observation is that the construction
we made in \S 2 is quite similar 
to a system of intersecting D-branes. 
In \S 2, we first introduced two lattice space-times and 
the matter fields come from the lattice variables living on links 
that connect them. 
Roughly speaking, we can regards the two lattice space-times 
as two bunches of D-branes and the link variables between them 
can be regarded as open strings between them.  
An important distinction is that we froze the degrees of freedom on 
the $N_f$-lattice by hand 
while it is automatic in intersecting D-branes. 
For example, in the case of a system of $N_c$ D1-branes 
and $N_f$ D5-branes, the gauge coupling constant on 
the D5-branes becomes effectively zero from the low energy 
effective theory point of view on the D1-branes 
since D5-brane is infinitely heavier than D1-brane.
The same thing might occur in the orbifold construction 
if, for example, 
we could change the dimensionality of the two 
lattice space-times. 
This would also be an important future work.

\section*{Acknowledgments}
The author would like to thank 
P.~H.~Damgaard, 
M.~Fukuma,
K.~Harada,
K.~Inoue,
S.~Iso,
I.~Kanamori,
D.~B.~Kaplan,
H.~Kawai,
K.~Murakami,
J.~Nishimura,
F.~Sugino
and
H.~Suzuki,
for useful discussion.
This work is supported by JSPS Postdoctoral Fellowship for Research Abroad.

\appendix

\section{Lattice formulation before the restriction (\ref{freeze})}

In this appendix, we write down the lattice action explicitly 
that is obtained by substituting (\ref{after projection 1}) 
and (\ref{after projection 2}) followed by the shift, 
$z_m \to 1/a + z_m$ and $\bz_m \to 1/a + \bz_m$.
This is a lattice formulation for a two-dimensional 
$U(N_c)\times U(N_f)$ quiver gauge theory 
coupled with matter fields in the bi-fundamental representation. 
When we considered matter fields in the fundamental representation, 
we regarded only the $N_c$-lattice in Fig.~\ref{lattice-fig}
as a ``real'' lattice space-time. 
In the case of the quiver gauge theory, however, 
we have to consider both of 
the $N_c$-lattice and the $N_f$-lattice. 
Then we express a position in the $N_c$-lattice by 
an integer valued two-vector $\bfk$
while the same position in the $N_f$-lattice is expressed 
by taking an underline as $\underline{\bfk}$. 
Correspondingly, it is convenient to change the notation 
of the matter fields since we have respected only the $N_c$-lattice 
in (\ref{after projection 2}). 
Instead of (\ref{after projection 2}), we use a new notation: 
\begin{align}
\phi &\equiv \sum_{\bfk\in\Z_N^2} \left(
-\sqrt{2}i \tbphi(\bfk,\underline{\bfk+\bfq})\otimes E_{\bfk,N+\bfk+\bfq}
+\sqrt{2}i \bphi(\underline{\bfk},\bfk+\bfq) \otimes E_{N+\bfk,\bfk+\bfq}
\right), \nn \\
\bphi &\equiv \sum_{\bfk\in\Z_N^2} \left(
-\sqrt{2}i \phi(\bfk+\bfq,\underline{\bfk})\otimes E_{\bfk+\bfq,N+\bfk}
+\sqrt{2}i \tphi(\underline{\bfk+\bfq},\bfk)\otimes E_{N+\bfk+\bfq,\bfk}
\right), \nn \\
\bareta &\equiv \sum_{\bfk\in\Z_N^2} \left(
\tbpsi(\bfk,\underline{\bfk+\bfq})\otimes E_{\bfk,N+\bfk+\bfq}
+\bpsi(\underline{\bfk},\bfk+\bfq)\otimes E_{N+\bfk,\bfk+\bfq}
\right), \nn \\
\bpsi_m&\equiv \sum_{\bfk\in\Z_N^2} \left(
\psi_m(\bfk+\hm+\bfq,\underline{\bfk})\otimes E_{\bfk+\hm+\bfq,N+\bfk}
+\tpsi_m(\underline{\bfk+\hm+\bfq},\bfk)\otimes E_{N+\bfk+\hm+\bfq,\bfk}
\right), \nn \\
\bxi_{mn}&\equiv \sum_{\bfk\in\Z_N^2} \left(
\tbpsi_{mn}(\bfk,\underline{\bfk+\hm+\hn+\bfq})\otimes E_{\bfk,N+\bfk+\hm+\hn+\bfq}
\right. \nn \\ &\left.
\hspace{5cm}
+\bpsi_{mn}(\underline{\bfk},\bfk+\hm+\hn+\bfq)\otimes E_{N+\bfk,\bfk+\hm+\hn+\bfq}
\right). 
\label{new matter notation}
\end{align}
The fields $\phi(\bfk,\underline{\bfk+\bfq})$, 
$\tbphi(\bfk+\bfq,\underline{\bfk})$, 
$\psi_m(\bfk+\hm+\bfq,\underline{\bfk})$, 
$\tbpsi(\bfk,\ul{\bfk+\bfq})$
and $\tbpsi_{mn}({\bfk},\ul{\bfk+\hm+\hn+\bfq})$ are in the 
representation $(\square,\overline{\square})$ of $U(N_c)\times U(N_f)$. 
Correspondingly, the field  $\phi(\bfk,\ul{\bfk+\bfq})$ lives on the 
link $(\bfk,\ul{\bfk+\bfq})$ and similar for the other fields. 
Similarly, 
$\bphi(\ul\bfk,\bfk+\bfq)$, 
$\tphi(\ul{\bfk+\bfq},\bfk)$, 
$\tpsi_m(\ul{\bfk+\hm+\bfq},\bfk)$, 
$\bpsi(\ul\bfk,\bfk+\bfq)$
and $\bpsi_{mn}(\ul\bfk,\bfk+\hm+\hn+\bfq)$ are in the 
representation $(\overline{\square},{\square})$. 
The filed $\bphi(\ul\bfk,\bfk+\bfq)$ 
lives on the link $(\ul\bfk,\bfk+\bfq)$ and so on. 

For a general field $\Phi(\bfk,\ul\bfl)$ 
in the representation $(\square,\overline{\square})$, 
we define covariant differences as 
\begin{align}
{\cD}_m^+ \Phi(\bfk,\ul\bfl) &\equiv 
 \nabla_m^+ \Phi(\bfk,\ul\bfl) 
 +z_m(\bfk)\Phi(\bfk+\hm,\ul{\bfl+\hm})-\Phi(\bfk,\ul\bfl)\hz_m(\ul\bfl), \nn \\
{\cD}_m^- \Phi(\bfk,\ul\bfl) &\equiv 
 \nabla_m^- \Phi(\bfk,\ul\bfl) 
 +z_m(\bfk-\hm)\Phi(\bfk,\ul\bfl)-\Phi(\bfk-\hm,\ul{\bfl-\hm})\hz_m(\ul{\bfl-\hm}),
 \nn \\
\bar{\cD}_m^+ \Phi(\bfk,\ul\bfl) &\equiv 
 \nabla_m^+ \Phi(\bfk,\ul\bfl) 
  +\Phi(\bfk+\hm,\ul{\bfl+\hm})\bhz_m(\bfl) 
  -\bz_m(\bfk)\Phi(\bfk,\ul{\bfl}),
 \nn \\
\bar{\cD}_m^- \Phi(\bfk,\bfl) &\equiv 
  +\Phi(\bfk,\ul{\bfl})\bhz_m(\bfl-\hm) 
  -\bz_m(\bfk-\hm)\Phi(\bfk-\hm,\ul{\bfl-\hm}). 
\end{align}
Similarly, for a general field $\bar{\Phi}(\ul\bfk,\bfl)$ 
in the representation $(\overline{\square},\square)$, 
we define 
\begin{align}
{\cD}_m^+ \bar\Phi(\ul\bfk,\bfl) &\equiv 
 \nabla_m^+ \bar\Phi(\ul\bfk,\bfl) 
 +\hz_m(\ul\bfk)\bar\Phi(\ul{\bfk+\hm},\bfl+\hm)
 -\bar\Phi(\ul{\bfk},\bfl)z_m(\bfl), \nn \\
{\cD}_m^- \bar\Phi(\bfk,\bfl) &\equiv 
 \nabla_m^- \bar\Phi(\bfk,\bfl) 
 +\hz_m(\ul{\bfk-\hm})\bar\Phi(\ul{\bfk},\bfl)
 -\bar\Phi(\ul{\bfk-\hm},\bfl-\hm)z_m(\bfl-\hm), \nn \\
\bar{\cD}_m^+ \bar\Phi(\ul\bfk,\bfl) &\equiv 
 \nabla_m^+ \bar\Phi(\ul\bfk,\bfl) 
 +\bar\Phi(\ul{\bfk+\hm},\bfl+\hm)\hz_m(\bfk)
 -\bhz_m(\ul\bfk)\bar\Phi(\ul\bfk,\bfl),
 \nn \\
\bar{\cD}_m^- \bar\Phi(\ul\bfk,\bfl) &\equiv 
 \nabla_m^- \bar\Phi(\ul\bfk,\bfl) 
 +\bar\Phi(\ul{\bfk},\bfl)\hz_m(\bfk-\hm)
 -\bhz_m(\ul{\bfk-\hm})\bar\Phi(\ul{\bfk-\hm},\bfl-\hm).
\end{align}

Using these notations, we can write down the action 
of the lattice theory as 
\begin{align}
 S&=S^{\rm boson} + S^{\rm fermion}, 
\end{align}
with
\begin{align}
 S^{\rm boson}=
{\rm Tr}_{N_c}\sum_{\bfk} \Bigl(
&\frac{1}{4}\left| \cF_{mn}(\bfk)\right|^2 
+\cD_m^+ \tbphi(\bfk,\ul{\bfk+\bfq}) 
\bar{\cD}_m^+ \tphi(\ul{\bfk+\bfq},\bfk) 
-\frac{1}{2} d^2(\bfk) \nn \\
&+\left[
\frac{1}{2}\cG(\bfk)
+\tbphi(\bfk,\ul{\bfk+\bfq})\tphi(\ul{\bfk+\bfq},\bfk)
-\phi(\bfk,\ul{\bfk-\bfq})\bphi(\ul{\bfk-\bfq},\bfk)
\right] d(\bfk)
\Bigr) \nn \\
+{\rm Tr}_{N_f}\sum_{\bfk}\Bigl(
&\frac{1}{4}\left| \hat\cF_{\bm\bn}(\ul\bfk)\right|^2 
+\cD_m^+ \bphi(\ul\bfk,{\bfk+\bfq}) 
\bar{\cD}_m^+ \tphi({\bfk+\bfq},\ul\bfk) 
-\frac{1}{2} \hd^2(\bfk) \nn \\
&+\left[
\frac{1}{2}\hat\cG(\ul\bfk)
+\bphi(\ul\bfk,{\bfk+\bfq})\phi({\bfk+\bfq},\ul\bfk)
-\tphi(\ul\bfk,{\bfk-\bfq})\tbphi({\bfk-\bfq},\ul\bfk)
\right] \hd(\bfk) 
\Bigr), \\
%%%%%%%%%%%%%%%%%%%%%%%%%%%%%%%%%%%%%%%%%%%%%%%%%%
S^{\rm fermion}=
{\rm Tr}_{N_c}\sum_{\bfk}\Bigl(
&-\lambda(\bfk)\bar\cD_m^-\lambda_m(\bfk)
+\frac{1}{2}\lambda_{mn}(\bfk)\left(
\cD_m^+\lambda_n(\bfk)-\cD_n^+\lambda_m(\bfk)
\right) \nn \\
&+\tbpsi(\bfk,\ul{\bfk+\bfq})\cD_m^-\tpsi_m(\ul{\bfk+\hm+\bfq},\bfk) \nn \\
&-\frac{1}{2}\tbpsi_{mn}(\bfk,\ul{\bfk+\hm+\hn+\bfq})\left(
\bar\cD_m^+\tpsi_n(\ul{\bfk+\hn+\bfq},\bfk)
-\bar\cD_n^+\tpsi_m(\ul{\bfk+\hm+\bfq},\bfk)
\right) \nn \\
&-\sqrt{2}i\lambda(\bfk)\bigl(
\phi(\bfk,\ul{\bfk-\bfq})\bpsi(\ul{\bfk-\bfq},\bfk)
-\tbpsi(\bfk,\ul{\bfk+\bfq})\tphi(\ul{\bfk+\bfq},\bfk)
\bigr) \nn \\
&+\sqrt{2}i\lambda_m(\bfk)\bigl(
\tbphi(\bfk+\hm,\ul{\bfk+\hm+\bfq})\tpsi_m(\ul{\bfk+\hm+\bfq},\bfk) \nn \\
&\hspace{5.2cm}-\psi_m(\bfk+\hm,\ul{\bfk-\bfq})\bphi(\ul{\bfk-\bfq},\bfk)
\bigr) \nn \\
&-\frac{\sqrt{2}i}{2}\lambda_{mn}(\bfk)\bigl(
\phi(\bfk,\ul{\bfk-\bfq})\bpsi_{mn}(\ul{\bfk-\bfq},\bfk+\hm+\hn) \nn \\
&\hspace{3.2cm}
-\tbpsi_{mn}(\bfk,\ul{\bfk+\hm+\hn+\bfq})
\tphi(\ul{\bfk+\hm+\hn+\bfq},\bfk+\hm+\hn)
\bigr) \Bigr) \nn \\
%%%%%%%%%%%%%%%%%%%%%%%%%%%%%%%
+{\rm Tr}_{N_f}\sum_{\bfk}\Bigl(
&-\hlambda(\ul\bfk)\bar\cD_m^-\hlambda_m(\ul\bfk)
+\frac{1}{2}\hlambda_{mn}(\ul\bfk)\left(
\cD_m^+\hlambda_n(\ul\bfk)-\cD_n^+\hlambda_m(\ul\bfk)
\right) \nn \\
&+\bpsi(\ul\bfk,{\bfk+\bfq})\cD_m^-\psi_m({\bfk+\hm+\bfq},\ul\bfk) \nn \\
&-\frac{1}{2}\bpsi_{mn}(\ul\bfk,{\bfk+\hm+\hn+\bfq})\left(
\bar\cD_m^+\psi_n({\bfk+\hn+\bfq},\ul\bfk)
-\bar\cD_n^+\psi_m({\bfk+\hm+\bfq},\ul\bfk)
\right) \nn \\
&+\sqrt{2}i\hlambda(\ul\bfk)\bigl(
\tphi(\ul\bfk,{\bfk-\bfq})\tbpsi({\bfk-\bfq},\ul\bfk)
-\bpsi(\ul\bfk,{\bfk+\bfq})\phi({\bfk+\bfq},\ul\bfk)
\bigr) \nn \\
&-\sqrt{2}i\hlambda_m(\ul\bfk)\bigl(
\bphi(\ul{\bfk+\hm},{\bfk+\hm+\bfq})\psi_m({\bfk+\hm+\bfq},\ul\bfk) \nn \\
&\hspace{5.2cm}-\tpsi_m(\ul{\bfk+\hm},{\bfk-\bfq})
\tbphi({\bfk-/\bfq},\ul\bfk)
\bigr) \nn \\
&+\frac{\sqrt{2}i}{2}\hlambda_{mn}(\ul\bfk)\bigl(
\tphi(\ul\bfk,{\bfk-\bfq})\tbpsi_{mn}({\bfk-\bfq},\ul{\bfk+\hm+\hn}) \nn \\
&\hspace{3.2cm}
-\bpsi_{mn}(\ul\bfk,{\bfk+\hm+\hn+\bfq})
\phi({\bfk+\hm+\hn+\bfq},\ul{\bfk+\hm+\hn})
\bigr) \Bigr), 
\end{align}
where $\cF_{mn}(\bfk)$, $\hat\cF_{mn}(\ul\bfk)$, 
$\cG(\bfk)$ and $\hat\cG(\ul(\bfk))$ 
are defined by 
\begin{align}
 \cF_{mn}(\bfk) &\equiv 
\nabla_m^+ z_n(\bfk) - \nabla_n^+ z_m(\bfk)
+z_m(\bfk)z_n(\bfk+\hm) - z_n(\bfk)z_m(\bfk+\hn), \nn \\
\hat\cF_{mn}(\ul\bfk) &\equiv 
\nabla_m^+ \hz_n(\ul\bfk) - \nabla_n^+ \hz_m(\ul\bfk)
+\hz_m(\ul\bfk)\hz_n(\ul{\bfk+\hm})
- \hz_n(\ul\bfk)\hz_m(\ul{\bfk+\hn}), \nn \\
\cG(\bfk) &\equiv \sum_m\left(
\nabla_m^- \left(z_m(\bfk)+\bz_m(\bfk)\right)
+z_m(\bfk)\bz_m(\bfk)-\bz_m(\bfk-\hm) z_m(\bfk-\hm)\right), \nn \\
\hat\cG(\ul\bfk) &\equiv \sum_m\left(
\nabla_m^- \left(\hz_m(\ul\bfk)+\bhz_m(\ul\bfk)\right)
+\hz_m(\ul\bfk)\bhz_m(\ul\bfk)
-\bhz_m(\ul{\bfk-\hm}) \hz_m(\ul{\bfk-\hm})\right).
\label{field strength}
\end{align}

\bibliographystyle{JHEP}
\bibliography{refs}

\end{document}